\documentclass{PoS}

\title{Progress in the calculation of purely leptonic $D_{(s)}$ -- decay constants from lattice QCD}

\ShortTitle{$D_{(s)}$ purely leptonic decays in lattice QCD}

\author{\speaker{Justus Tobias Tsang}\\
  Higgs Centre for Theoretical Physics\\
  School of Physics \& Astronomy, University of Edinburgh\\
  EH9 3JZ, Edinburgh, United Kingdom\\
  E-mail: \email{j.t.tsang@ed.ac.uk}}

\abstract{We review recent progress in the calculation of the decay constants $f_{D}$ and $f_{D_s}$ from lattice QCD. We focus particularly on simulations with $N_f=2+1$ and $N_f=2+1+1$ and simulations with close to physical light quark masses.}

\FullConference{9th International Workshop on the CKM Unitarity Triangle\\
		28  November - 3 December 2016\\
		Tata Institute for Fundamental Research (TIFR), Mumbai, India}

\usepackage[square,numbers,sort&compress]{natbib}            
\newcommand{\abs}[1]{\left| #1 \right|}                                         
\newcommand{\eq}[1]{\begin{equation} #1 \end{equation}}
\newcommand{\matrixel}[3]{\left< #1 \vphantom{#2#3} \right| #2 \left| #3 \vphantom{#1#2} \right>}   
\usepackage{multirow}
\usepackage{tablefootnote}

\begin{document}

\section{Introduction}\label{sec:intro}
In the Standard Model (SM) the Cabibbo-Kobayashi-Maskawa (CKM) matrix is unitary. To test the unitarity of the second row of the CKM matrix requires the determination of the matrix elements $\abs{V_{cd}}$ and $\abs{V_{cs}}$.
Neglecting electromagnetic corrections, the SM decay rates of $D$ and $D_s$ mesons can be factorised into these CKM matrix elements, the decay constants $f_D$ and $f_{D_s}$ and known kinematic factors. Experimental measurements of the decay rates combined with non-perturbative calculations of $f_{D_{(s)}}$ from lattice QCD calculations hence allow for a determination of $\abs{V_{cd}}$ and $\abs{V_{cs}}$.
The current global averages of $\abs{V_{cq}}f_{D^+_{(s)}}$ are~\cite{Olive:2016xmw}
\eq{
  \abs{V_{cd}}f_{D^+} = 45.91 \pm 1.05\,\mathrm{MeV} \qquad \mathrm{and} \qquad \abs{V_{cs}}f_{D^+_s} = 250.9\pm 4.0\,\mathrm{MeV}.
  \label{eq:exp}
}

On the lattice, the decay constants $f_{D_{(s)}}$ are extracted from the Euclidean matrix elements
\eq{
  \matrixel{0}{\overline{c}\gamma_5\gamma^\mu q}{P(p)} = i f_P\, p^\mu_{P}
}
for the pseudoscalar mesons $P = D, D_s$ and $q=d, s$, respectively. 

Lattice QCD calculations differ in many aspects: e.g. in the number of flavours $N_f$ in the sea, the lattice gauge and fermion actions, the used lattice spacings and how these are determined, and the simulated pion masses. In all cases an inter- or extrapolation to physical quark masses (chiral inter/extrapolation for the light quark mass) as well as an extrapolation to vanishing lattice spacing (continuum limit) is required and only results which carried out these extrapolations will be considered.

In section \ref{sec:results} we will review the recent progress in the determination of $f_{D_{(s)}}$ from lattice QCD calculations. Finally, in section \ref{sec:outlook} we will conclude and give an outlook of current and ongoing calculations.
\section{Lattice results}\label{sec:results}
\begin{figure}
  \begin{center}
    \includegraphics[width=.49\textwidth]{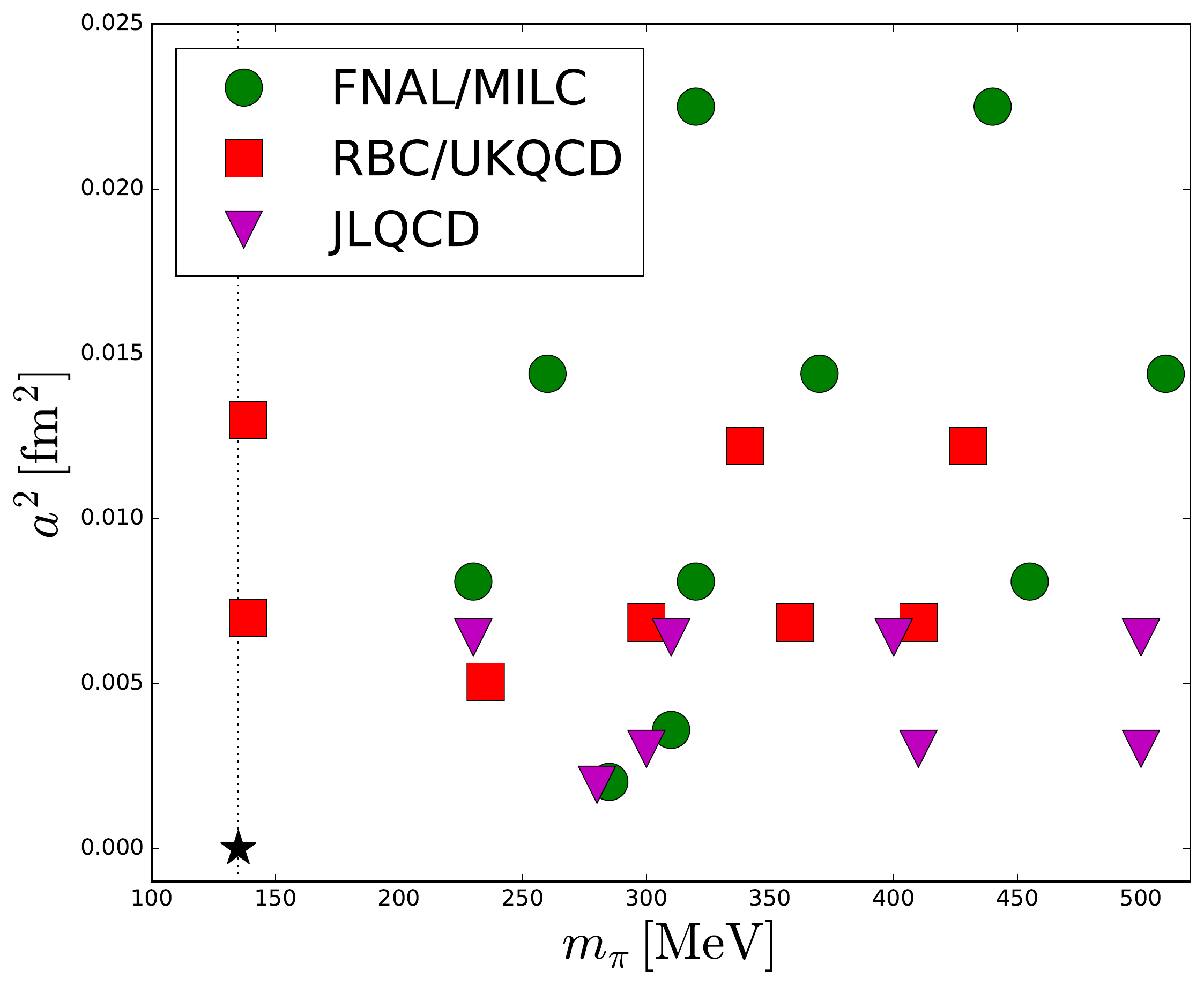}
    \includegraphics[width=.49\textwidth]{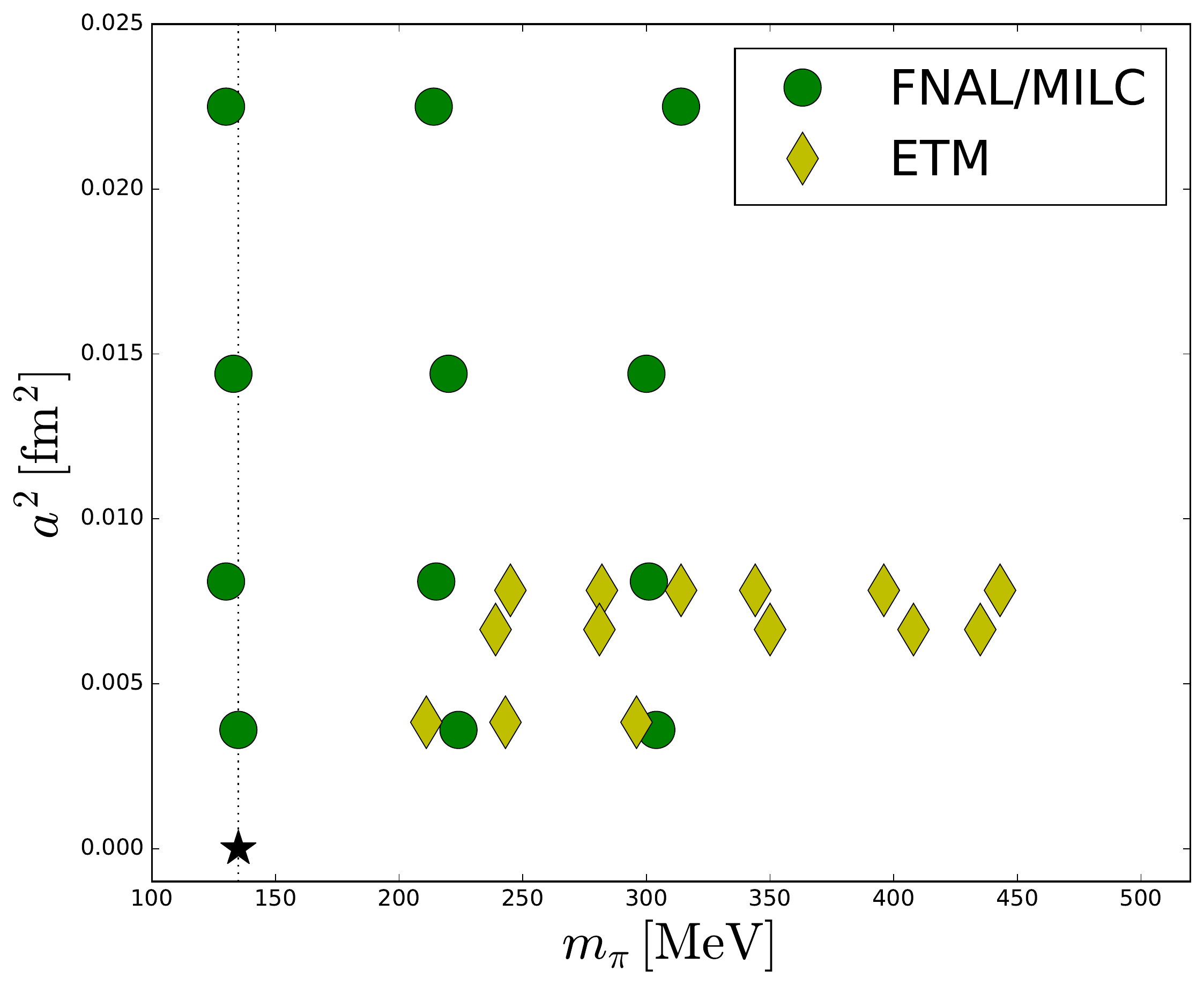}
  \end{center}
  \caption{The $N_f=2+1$ (left) and $N_f=2+1+1$ (right) flavour ensembles which have recently been used for the calculation of $D_{(s)}$ meson decay constants. The vertical black dotted line corresponds to physical pion masses, the black star denotes the physical point, to which simulated data needs to be extrapolated.}
  \label{fig:ensembles}
\end{figure}

Most modern simulations use two degenerate light quarks and a strange quark in the sea sector ($N_f=2+1$ flavour simulations)~\cite{Boyle:2017jwu,Fahy:2017enl,Yang:2014sea,Na:2012iu,Davies:2010ip,Bazavov:2011aa}, but first results including a dynamical charm quark ($N_f=2+1+1$ flavours) are available~\cite{Bazavov:2014wgs,Carrasco:2014poa}. Details of the ensembles used for these works are shown in figure \ref{fig:ensembles}. Results which appeared before 30 November 2015 are summarised in the most recent FLAG review~\cite{Aoki:2016frl}. A summary including more recent results is given in table \ref{tab:fDresults}. All these results are in good agreement with each other.

Currently there are only two published results with $N_f=2+1+1$ flavours in the sea sector~\cite{Bazavov:2014wgs,Carrasco:2014poa} and only two results which use ensembles with (near) physical pion masses~\cite{Boyle:2017jwu,Bazavov:2014wgs}.

The most precise prediction to date is obtained by the FNAL/MILC collaborations~\cite{Bazavov:2014wgs} using $N_f=2+1+1$ flavours of highly improved staggered quarks (HISQ) in the sea and the valence sector, four lattice spacings and including ensembles with physical pion masses. The dominant systematic error in this work arises from the combined chiral and continuum limit extrapolation.

The ETM collaboration has also published a result with $N_f=2+1+1$ flavours~\cite{Carrasco:2014poa} of twisted mass (TM) fermions in the sea sector. In the valence sector they use TM fermions for the light quarks and Osterwalder-Seiler fermions for the strange and the charm quark, resulting in a mixed action setup. They first take the chiral continuum limit of $f_{D_s}/m_{D_s}$ before extracting $f_{D_s}$ in the continuum. Similarly, the double ratio $(f_{D_s}/f_D)/(f_K/f_\pi)$ is extrapolated to the physical point before extracting $f_{D_s}/f_D$ and finally $f_D$ from this. The dominant quoted error is an accumulation of the scale setting, statistics and all errors of inputs needed to eventually extract $f_D$.

\begin{table}
  \begin{center}
    \resizebox{\textwidth}{!}{
      \begin{tabular}{|c|r|ll||ccc|}
        \hline
        $N_f$ & Collaboration & $f_{D}\,[\mathrm{MeV}]$ & $f_{D_s}\,[\mathrm{MeV}]$ & $m_\pi^\mathrm{min}\,[\mathrm{MeV}]$ & $a\,[\mathrm{fm}]$ & \# $a$\\
        \hline\hline
        \multirow{6}{*}{$2+1$} & RBC/UKQCD~\cite{Boyle:2017jwu}  & $208.7(2.8)\left(^{+2.1}_{-1.8}\right)$ & $246.4(1.3)\left(^{+1.3}_{-1.9}\right)$ & {\bf 139} & 0.07-0.11 & 3\\
        & JLQCD~\cite{Fahy:2017enl}\tablefootnote{No full systematic error budget has been published yet. The second quoted error arises from the scale setting only.} & 212.8(1.7)(3.6) & 244.0(0.8)(4.1) & 230 & 0.044-0.08 & 3\\
        & $\chi$QCD~\cite{Yang:2014sea}       &                 & 254(2)(4)       & 300 & 0.08-0.11  & 2\\
        & HPQCD~\cite{Na:2012iu}              & 208.3(1.0)(3.3) & 246.0(0.7)(3.5) & 245 & 0.08-0.12  & 2\\
        & HPQCD~\cite{Davies:2010ip}          &                 & 248.0(2.5)      & 260 & 0.045-0.15 & 5\\
        & FNAL/MILC~\cite{Bazavov:2011aa}     & 218.9(11.3)     & 260.1(10.8)     & 230 & 0.09-0.15  & 3\\
        \hline
        \multirow{2}{*}{$2+1+1$} & FNAL/MILC~\cite{Bazavov:2014wgs} & $212.6(0.4)\left(^{+1.0}_{-1.2}\right)$ & $249.0(0.3)\left(^{+1.1}_{-1.5}\right)$ & {\bf 130} & 0.06-0.15 & 4\\
        & ETM~\cite{Carrasco:2014poa} & $207.4(3.8)$ & 247.2(4.1) & 210 & 0.06-0.09 & 3\\
        \hline
      \end{tabular}
    }
  \end{center}
  \caption{Summary of results for $f_D$ and $f_{D_s}$ with $N_f=2+1$ and $N_f=2+1+1$ flavours in the sea sector. Where multiple results exist within the same setup, we only quote the most precise value. Where two errors are given, they correspond to the statistic and systematic errors, respectively. The only exception is the JLQCD result, where no full systematic error budget has been published yet. The second error quoted here arises from the scale setting only.
    The smallest value of the pion mass, that enters the calculations is denoted by $m_\pi^\mathrm{min}$, $a$ gives the range of lattice spacings and $\# a$ the number of distinct lattice spacings used in the calculation. }
  \label{tab:fDresults}
\end{table}

Even though table \ref{tab:fDresults} lists 6 different results for simulations with $N_f=2+1$, these are based on only three distinct sets of gauge ensembles. In particular the results of refs \cite{Na:2012iu,Davies:2010ip,Bazavov:2011aa} are based on overlapping subsets of the FNAL/MILC gauge ensembles using asqtad rooted staggered quarks. 
In 2010 HPQCD~\cite{Davies:2010ip} used HISQ valence quarks to obtain a prediction for $f_{D_s}$ on these ensembles. Their dominant errors arise from the absolute scale setting (0.57\%) which is of the same size as the combined error due to statistics and valence tuning.
In 2011, the FNAL/MILC collaborations~\cite{Bazavov:2011aa} used the Fermilab action for the charm quark, leading to comparably large heavy quark discretisation errors, which constitutes their dominant contribution to the final error and which is estimated to be $8.2\,\mathrm{MeV}$ and $8.3\,\mathrm{MeV}$ for $f_D$ and $f_{D_s}$, respectively.
Finally, in 2012 HPQCD made a prediction for $f_D$ based on HISQ valence quarks on two lattice spacings~\cite{Na:2012iu}. The leading systematic error here arises from need of a chiral extrapolation to physical pion masses.

Refs \cite{Boyle:2017jwu,Yang:2014sea} use RBC/UKQCD's gauge ensembles with domain wall fermions. In 2015 the $\chi$QCD collaboration~\cite{Yang:2014sea} published a prediction for $f_{D_s}$ by using Overlap valence quarks. They quote three leading contributions to the total error that are of nearly similar size (0.8-1.0\%): These are the statistical error (which includes discretisation errors and the chiral extrapolation), the quark mass dependence of the scale setting procedure and the limited reach in the heavy quark mass on the coarser set of ensembles.
A new result by the RBC/UKQCD collaborations appeared in 2017~\cite{Boyle:2017jwu}. This result includes three lattice spacings as well as two ensembles with physical pion masses. Domain wall fermions are used for all quarks. Whilst the parameters of the domain wall action for valence light and strange quarks are kept the same as in the sea, the charm quarks are simulated with a slight modification resulting in reduced lattice artefacts~\cite{Cho:2015ffa,Boyle:2016imm}. The leading systematic error is quantified using variations of the global fit, which simultaneously carries out the chiral and continuum limit extrapolation and the interpolation to the physical charm quark mass.

The JLQCD collaboration~\cite{Fahy:2017enl} recently presented results using stout smeared domain wall fermions in the sea and the valence sector. However, to date, only the systematic error due to the scale setting has been discussed.

The RQCD and ALPHA collaborations have recently presented a status update of a new calculation~\cite{Collins:2017iud} with $N_f=2+1$ flavours of non-perturbatively improved Wilson quarks using pion masses as low as $200\,\mathrm{MeV}$ and currently including two lattice spacings with $a=0.064,0.085\,\mathrm{fm}$. RBC/UKQCD also presented results with a choice of domain wall parameters that allows for further reach in the heavy quark mass~\cite{Boyle:2016lzk}. 

To significantly improve lattice QCD predictions for $f_{D_{(s)}}$ beyond the percent level, electromagnetic corrections and isospin breaking will have to be considered.
FNAL/MILC~\cite{Bazavov:2014wgs} presented a first estimate of the isospin breaking effect on the decay constant $f_D$. By considering valence light quark masses $m_l=(m_u+m_d)/2$ and $m_l=m_d$ they found $f_{D^+}-f_D = 0.47(1)_\mathrm{stat}\left(^{+25}_{-\hphantom{0}6}\right)_\mathrm{sys}\,\mathrm{MeV}$.

\section{Conclusions and outlook}\label{sec:outlook}
A number of different lattice calculations with $N_f=2+1$ and $N_f=2+1+1$ flavours have resulted in predictions for $f_D$ and $f_{D_s}$ which have started to reach percent precision. Used fermion actions include asqtad, HISQ, twisted mass and domain wall fermions in the sea. In the valence sector there is even more variation additionally including overlap, Fermilab and Osterwalder-Seiler fermions. However, the number of distinct underlying gauge configurations is still limited, as are calculations with $N_f=2+1+1$ flavours and calculations with physical pion masses. Given that the dominant systematic error of most calculations arises from the chiral and continuum limit extrapolations, the use of physical pion masses and finer gauge ensembles will lead to improvements over the current precision.

Currently, the uncertainty on the extraction of the corresponding CKM matrix elements is limited by the experimental precision of 2.3\% and 1.6\%, respectively (cf (\ref{eq:exp})), so more experimental data is needed to constrain $\abs{V_{cd}}$ and $\abs{V_{cs}}$ more tightly.

\acknowledgments{JTT is supported by STFC, grant ST/L000458/1.}
    {\small
      \bibliographystyle{JHEP}
      \bibliography{CKM.bib}
}

\end{document}